\documentclass[aip,apl,numerical,reprint]{revtex4-1}
\usepackage{graphicx}
\usepackage{upgreek}
\newcommand{\micro}{${\upmu}$}

\begin{document}

\title{Strong exciton-photon coupling in open semiconductor microcavities} 

\author{S. Dufferwiel} 
\author{F. Fras}
\affiliation{Department of Physics and Astronomy, University of Sheffield, S3 7RH, UK}
\author{A. Trichet}
\affiliation{Department of Materials, University of Oxford, OX1 3PH, UK}
\author{P. M. Walker}
\author{F. Li}
\author{L. Giriunas}
\author{M. N. Makhonin}
\affiliation{Department of Physics and Astronomy, University of Sheffield, S3 7RH, UK}
\author{L. R. Wilson}
\affiliation{Department of Physics and Astronomy, University of Sheffield, S3 7RH, UK}
\author{J. M. Smith}
\affiliation{Department of Materials, University of Oxford, OX1 3PH, UK}
\author{E. Clarke}
\affiliation{\mbox{EPSRC National Centre for III-V Technologies, University of Sheffield, S1 3JD, UK}}
\author{M. S. Skolnick}
\author{D. N. Krizhanovskii}
\affiliation{Department of Physics and Astronomy, University of Sheffield, S3 7RH, UK}

\date{\today}

\begin{abstract}
We present a method to implement 3-dimensional polariton confinement with in-situ spectral tuning of the cavity mode. Our tunable microcavity is a hybrid system consisting of a bottom semiconductor distributed Bragg reflector (DBR) with a cavity containing quantum wells (QWs) grown on top and a dielectric concave DBR separated by a micrometer sized gap. Nanopositioners allow independent positioning of the two mirrors and the cavity mode energy can be tuned by controlling the distance between them. When close to resonance we observe a characteristic anticrossing between the cavity modes and the QW exciton demonstrating strong coupling. For the smallest radii of curvature concave mirrors of 5.6 \micro m and 7.5 \micro m real-space polariton imaging reveals submicron polariton confinement due to the hemispherical cavity geometry.
\end{abstract}

\pacs{}

\maketitle 
Strong coupling between quantum well (QW) excitons and photons in a semiconductor microcavity leads to the formation of quasiparticles known as exciton-polaritons.\cite{Weisbuch1992} Due to their part-matter part-photon nature, polaritons possess a large nonlinearity arising from exciton-exciton interactions that can be probed through coupling to their photonic component. Polaritons exhibit interesting phenomena such as non-equilibrium condensation\cite{Kasprzak2006}, superfluid like behaviour\cite{Amo2009} and soliton formation\cite{Sich2012}. Conventional semiconductor microcavities are comprised of two semiconductor distributed Bragg reflectors (DBRs) separated by a $m\lambda/2$ cavity containing one or more QWs located at electric-field antinodes. In these monolithic cavities the spectral features are fixed in energy and in-situ tuning of the cavity mode is limited to tapering of the cavity length along the sample.\cite{Weisbuch1992} In an open cavity system the microcavity is split into three separate parts; a bottom DBR with cavity region, an air gap and a top DBR. The cavity resonance can then simply be tuned by changing the distance between the mirrors. Open cavity systems have been used to study cavity quantum electrodynamical effects with a variety of emitters.\cite{Colombe2007,Hunger2010,Muller2010,MiguelSanchez2013,Barbour2011,Ziyun2012} 

Due to the low in plane effective mass of polaritons micrometer sized lateral confinement leads to the discretisation of the polariton modes. Smaller confinement causes the average polariton-polariton distance to be reduced and the nonlinear exciton interaction to increase for the same excitation power. Combined with narrow polariton linewidths this lateral confinement may lead to the currently unobserved effect of polariton blockade.\cite{Verger2006} Several systems have been developed to achieve zero-dimensional polaritons based on excitonic or photonic confinement. Spatial modulation of the polariton potential has been introduced using cross propagating surface acoustic waves (SAWs) but the confinement size is limited by the small penetration into the microcavity at high SAW frequencies.\cite{CerdaMendez2010,cerdamendez2013}  The application of controlled stress can be used to spatially localise the exciton wavefunction but the energy spectrum remains a quasi-continuum due to weak confinement.\cite{Negoita1999,Balili2007} Alternatively photonic confinement can be introduced through the growth of mesa structures using patterned regrowth\cite{Daif2006}, post growth etching into micropillars \cite{Ferrier2011} and photonic crystals\cite{Azzini2011}. Typically micropillars and photonic crystal microcavities suffer from increased losses as the dimensions are reduced and in all mentioned cases the spectral tuning of the cavities is limited. 

\begin{figure}
\center
\includegraphics{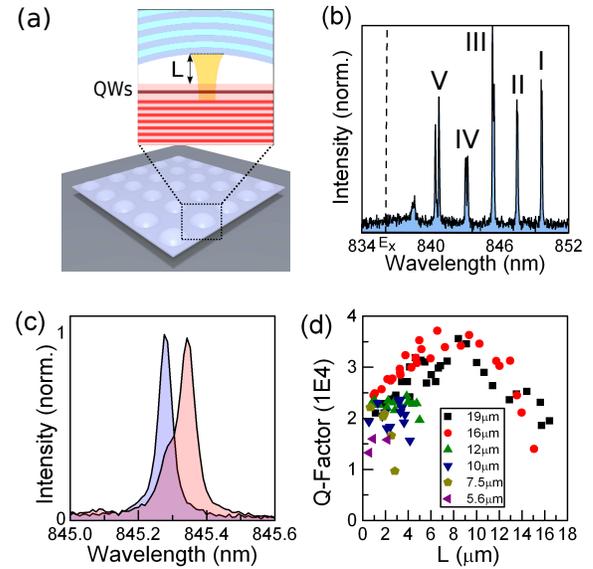}
\caption{\label{fig1} (a) The open cavity system formed by an array of dielectric concave DBRs and a semiconductor DBR containing QWs. The inset shows the formed hemispherical cavity with the mirrors separated by a small gap.  (b) Typical photoluminescence (PL) spectrum at large negative exciton-photon detuning showing the formation of a ground longitudinal mode and higher order transverse modes due to lateral confinement (c) The ground longitudinal mode at large negative detuning shows an orthogonally polarised splitting of 110 \micro eV due to birefringence. (d) Photonic Q-factor as a function of mirror separation for concave mirror RoCs of 19 \micro m, 16 \micro m, 12 \micro m, 10 \micro m, 7.5 \micro m and 5.6 \micro m.}
\end{figure}

In this letter we present a fully tunable zero dimensional polariton system based upon strong photonic confinement in an open hemispherical microcavity.\cite{Dolan2010} The tuneable cavity consists of a dielectric DBR containing concave features of various radii of curvature (RoC) and a semiconductor DBR with a cavity region containing QWs grown above. The hemispherical cavity geometry leads to strong lateral photonic confinement which is then imprinted in the polariton wavefunction. The combination of both longitudinal and lateral confinement creates photonic confinement in 3-dimensions and leads to the formation of 0-dimensional polaritons. 

The dielectric DBR contains an array of concave mirrors fabricated using focused ion beam (FIB) milling before coating with dielectric layers.\cite{Dolan2010} In contrast to fiber based open microcavities, where the concave feature is fabricated using laser ablation into the end of an optical fiber\cite{Colombe2007,Hunger2010,Muller2010,MiguelSanchez2013} FIB milling on a planar substrate allows both the fabrication of smaller RoCs and allows a single sample to contain an array of concave mirrors. We note that laser ablation has also been used to form concave depressions in a planar substrate to fabricate a similar cavity to that described here, rather than using a fiber based approach.\cite{Barbour2011} Fig.~\ref{fig1}(a) shows the set-up of the open cavity system consisting of a top dielectric concave DBR array and a bottom semiconductor DBR with a cavity region containing QWs that are separated by a micron sized gap. The two mirrors were mounted on to two separate xyz-piezo closed loop attocube nanopositioner stacks allowing full spatial and spectral tuning of the cavity. In this study we report on experiments using a semiconductor MBE grown 27 paired Al$_{0.85}$Ga$_{0.15}$As/GaAs DBR with a cavity region grown on top consisting of two sets of three 10 nm InGaAs QWs placed at E-field antinodes using GaAs spacer layers. The dielectric concave mirrors have RoCs of 19 \micro m, 16 \micro m, 12 \micro m, 10 \micro m, 7.5 \micro m and 5.6 \micro m with the desired RoC mirror placed into the optical path using the xy-stages. For the semiconductor mirror the piezo stack also consists of two tilt goniometer stages allowing full control of the parallelism between the mirrors. For low temperature measurements at 4K the homemade system is placed in a bath cryostat with a small amount of He exchange gas and is securely fastened inside a liquid He dewar. Optical access to the samples is provided via an optical table placed on top of the insert with free space access through an optical window.

\begin{figure}
\center
\includegraphics{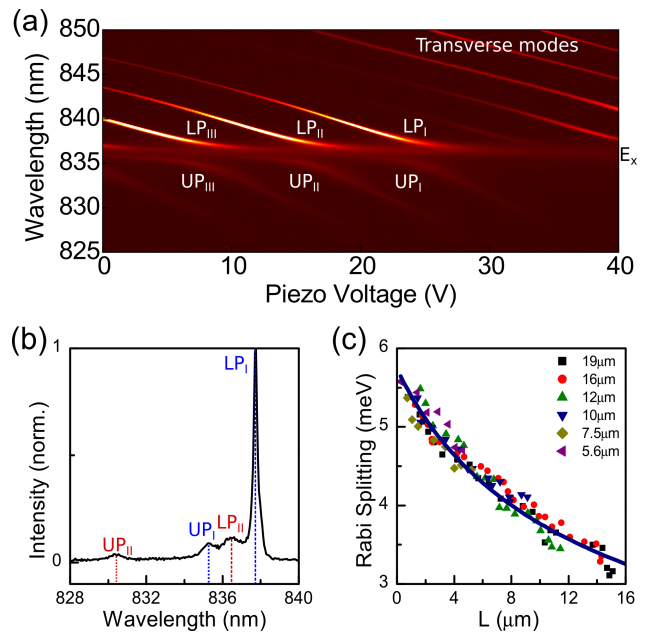}
\caption{\label{fig2} (a) Typical avoided crossings as the cavity modes are tuned through resonance with the QW exciton. The RoC of the concave mirror was 19 \micro m and the mirror separation decreases with increasing piezo voltage.  As well as the longitudinal mode we observe strong coupling between the exciton and all observed higher order transverse modes. (b) Spectral slice close to resonance between the ground mode I and QW exciton. The blue dashed and dotted lines indicates the lower ($LP_I$) and upper ($UP_I$) polariton arising from exciton coupling with mode I with a splitting of 4.4 meV at zero detuning. The red indicates the lower ($LP_{II}$) and upper ($UP_{II}$) polariton branches arising from transverse mode II. The mirror separation $\mathit{L}$ $\sim$ 5 \micro m. (c) Rabi splitting at zero exciton-photon detuning as a function of mirror separation for concave mirror RoCs of 19 \micro m, 16 \micro m, 12 \micro m, 10 \micro m, 7.5 \micro m and 5.6 \micro m.}
\end{figure}

For a hemispherical cavity, stable modes are only formed when the stability condition $L_{phys} \le$ RoC is satisfied,\cite{siegman86} where $\mathit{L_{phys}}$ is the physical cavity length and must take into account the field penetration into the DBRs. 
For our composite cavity the physical length is defined by \begin{equation} L_{phys} = L + L_{DBR 1} + L_{DBR 2} +L_{QW} \end{equation} where $\mathit{L}$ is the tunable mirror separation distance as indicated in Fig.~\ref{fig1}(a), $\mathit{L_{DBR 1}}$ and $\mathit{L_{DBR 2}}$ are the physical field penetration depths into the dielectric and semiconductor DBRs and $\mathit{L_{QW}}$ is the physical length of the cavity-QW containing region. 
The longitudinal spectral resonances of the cavity are determined by the condition that the round trip phase in the cavity ${\phi}(k,L)$ is an integer multiple of 2${\pi}$. \begin{equation} {\phi}(k,L) = 2kL + \phi_{DBR}(k) = 2m{\pi}\end{equation} where $\mathit{m}$ is an integer, $\mathit{k}$ is the vacuum wavenumber and $\mathit{L}$ is the mirror separation distance. The structural constant ${\phi}_{DBR}(k)$ is the sum of the reflection phases of the top DBR and combined bottom DBR and cavity region and may be calculated using a transfer matrix technique. 
This formula may be rearranged to allow extraction of the mirror separation $L$ from the free spectral range $\Delta \lambda = {\lambda}_{(m-1)}- {\lambda}_{(m)}$ between adjacent longitudinal modes. The usual expression becomes modified to account for the difference in DBR reflection phase ${\Delta}{\phi}_{DBR} = {\phi}_{DBR,(m-1)} - {\phi}_{DBR,(m)}$ at the two wavelengths of the adjacent longitudinal modes. \begin{equation} L = \frac{{\lambda}_{m}^2}{2{\Delta} {\lambda}} \left(1 + {\Delta}{\phi}_{DBR} / 2{\pi} \right) \label{L_FSR}\end{equation} 

Fig.~\ref{fig1}(b) shows a typical photoluminescence (PL) spectrum with the longitudinal cavity mode at a large negative detuning from the exciton. In this regime the lower polariton is largely photonic and the optical characteristics of the cavity can be probed. We use non-resonant excitation close to a cavity reflectivity minimum using a 685 nm laser diode. Spectroscopy is performed using a 0.75 m monochromator with a cooled CCD at $-70^{\circ}C$. In addition to the ground longitudinal cavity mode I we observe a number of higher order transverse modes (labeled II, III, IV, V) with equal energy spacing due to lateral confinement. This suggests that the transverse photonic potential created by the curved top mirror is nearly parabolic. Fig.~\ref{fig1}(c) shows that for a concave mirror of 19 \micro m mode I exhibits a splitting between orthogonally polarised modes of $\sim$110 \micro eV, which probably arises from birefringence in the bottom and/or in the top mirror due to stress. Higher order transverse modes also split into a doublet with an energy splittings of the order of $\sim$100-200 \micro eV as seen in Fig.~\ref{fig1}(b) due to a combination of birefringence and breaking of the cylindrical symmetry in the shape of the top mirror as discussed below.

Fig.~\ref{fig1}(d) shows the photonic Q-factor as a function of $L$ measured for mirrors with different RoC, where L was deduced from white light reflectivity spectra using Eqn.~\ref{L_FSR}. The photonic Q-factor of the microcavity increases with cavity length due to the increased photonic lifetime up to a maximum value of $\sim35,000$ at $\mathit{L} =$ 9 \micro m for the mirror with RoC = 19  \micro m and 16 \micro m. The Q-factor then decreases as $L$ and hence $L_{phys}$ increase further. This can be attributed to the diverging beam waist on the concave mirror leading to larger losses as we approach the limits of the stability condition.\cite{Dolan2010} We estimate that the minimum mirror separation before touching that we can reach is $\mathit{L}$ $\approx$ 1 \micro m.

\begin{figure}
\center
\includegraphics{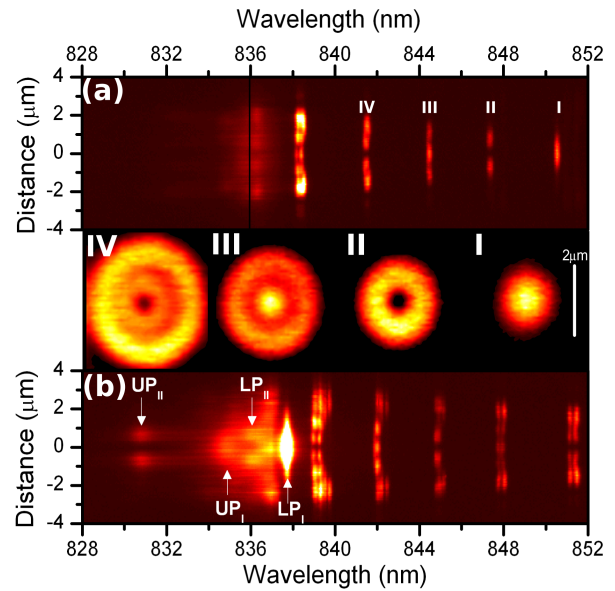}
\caption{\label{fig3} (a) Position-wavelength images of cavity modes when longitudinal mode I is at a very large negative exciton-photon detuning of around -5$\Omega_{Rabi}$ = -26 meV. The black vertical line indicates the QW exciton energy. The concave mirror RoC was 19 \micro m. Insets I-IV: Real space PL images of photonic modes revealing profiles for modes I, II, III and IV. (b) Position-wavelength images of cavity modes when ground mode I is close to resonance with the QW exciton. The formation of both $UP_{I}$ /$LP_{I}$ at resonance and $UP_{II}$/$LP_{II}$ at a positive exciton-photon detuning of approximately  1.5$\Omega_{Rabi}$ = 7.8 meV are labelled. At longer wavelengths we see a number of high order transverse modes associated with another longitudinal mode at lower energy.}
\end{figure}

In order to spectrally tune the cavity resonance we apply a DC voltage to the bottom z-piezo nanopositioner which decreases the mirror separation $\mathit{L}$. By scanning the cavity length in this manner we tune the cavity modes through resonance with the QW exciton energy. Fig.~\ref{fig2}(a) shows the characteristic avoided crossing in PL between the cavity modes and QW exciton. It is clear that both mode I and higher order transverse modes all display an avoided crossing with the exciton resonance. The Rabi splitting $\Omega_{Rabi}$, when mode I is at resonance with the QW exciton at a mirror separation $\mathit{L}$ $\sim$ 5 \micro m, is 4.4 meV and is comparable across all modes due to the negligible dependence of the coupling strength with exciton wavevector. 
Fig.~\ref{fig2}(b) shows a spectral slice at zero detuning between mode I and the QW exciton. The confined upper polariton has a weaker PL signal as polaritons tend to relax towards the lowest energy states of the trap. The upper polariton is significantly broader than the lower polariton due to scattering with phonons and QW disorder potential to the lower polariton and the exciton reservoir.\cite{Savona1997}
At zero exciton-photon detuning at the minimum effective cavity length we measure a lower polariton linewidth of 0.26 meV. The bare QW exciton linewidth is 1.2 meV and the photonic linewidth is around 78 \micro eV. At the minimum cavity length the maximum observed vacuum Rabi splitting at resonance is 5.6 meV which is comparable to the values obtained in monolithic 6 QW $3\lambda/2$ microcavities. In Fig.~\ref{fig2}(c) we plot the Rabi splitting at zero exciton-photon detuning as a function of $L$ for cavities formed with each of the concave RoCs. The splitting is expected to be inversely proportional to the square root of the cavity effective length $L_{eff}$, which is given by the ratio of the integrated electric energy density in the cavity divided by the density at the QWs.\cite{Kavokin2007}  At fixed energy $L_{eff}$ is proportional to $L + C$ where $\mathit{L}$ is the mirror separation and the constant $\mathit{C}$ accounts for the fraction of the mode energy located in the DBRs and QW region. The fit in Fig.~\ref{fig2}(c) corresponds to $\Omega_{Rabi} \propto 1/\sqrt{L + C}$ with $C \approx 7.6$ \micro m. No dependence on the RoC was observed.

To enable imaging of the polariton modes we employ a wound fiber bundle consisting of a 4$\times$4 mm array of single mode fibers. Each single mode fiber in the array acts like a single pixel allowing the image focused on one end of the fiber to be emitted from the other end. We then image the real or $k$--space image of the polariton on to one end facet and image the other end on to the slits of a spectrometer.  

Fig.~\ref{fig3}(a) shows position-wavelength images of cavity modes when longitudinal mode I is negatively detuned with respect to the exciton. Here the emission intensities are plotted for different wavelengths versus position across the line going through the middle of the cavity. These spatial mode profiles correspond to the modes that we display spectrally in Fig.~\ref{fig1}(b) and provide direct evidence of the micrometric sized confinement and spatial discretization of modes. Fig.~\ref{fig3}(b) shows the position-wavelength images of polariton modes when the mode I is close to resonance with the QW exciton. Here we can clearly see the imprinting of the photonic spatial distribution into the polariton modes $UP_{I}$ and $LP_{I}$ and $UP_{II}$ and $LP_{II}$. Modes $UP_{I}$ and $LP_{I}$ are characterised by Gaussian spatial distribution, whereas modes $UP_{II}$ and $LP_{II}$ have two distinct maxima at ~$\pm1$ \micro m. For the longitudinal mode I we measure a Gaussian beam waist size of 1.16 \micro m on the concave mirror with RoC = 19 \micro m at $\mathit{L} \approx$ 1.7 \micro m. For the RoCs of 7.5 \micro m and 5.6 \micro m we measure beam waist sizes of 0.85 \micro m and 0.78 \micro m -- better than the confinement achieved in mesa\cite{Daif2006}or micropillar\cite{Ferrier2011} structures. We note that the beam waist size on the planar semiconductor part is slightly smaller than the beam waist on the concave mirror when $L_{phys} \ll$ RoC.\cite{siegman86} 

The inset in Fig.~\ref{fig3} shows the real space PL images of photonic modes revealing profiles for modes I, II, III and IV, which resemble helical Laguerre-Gaussian (LG) transverse modes. Formation of such modes is expected in a system with perfect cylindrical symmetry.\cite{Meucci1997} Nevertheless, spectrally resolved images reveal that LG in our system actually are not the eigenstates and are split into a family of Mathieu and Ince-Gaussian modes\cite{Nardin2010} probably due to breaking of the cylindrical symmetry of the top mirror. The detailed investigation of this effect is beyond the scope of the present manuscript and will be presented elsewhere.  

In summary we present a novel system for strong polariton confinement with in situ tuning of the cavity modes. Combining the submicron confinement with very narrow lower polariton linewidths has the potential to lead to the manifestation of strong nonlinear interactions between polaritons and the observation of the polariton blockade effect.\cite{Verger2006} A significant reduction in the polariton linewidth can be expected by using a QW sample with less inhomogeneous broadening. In addition the versatile fabrication method will allow the production of polariton lattices or molecules where the effect of polariton blockade is predicted to be enhanced by an order of magnitude.\cite{Bamba2011} Fabrication of 2-dimensional arrays of concave mirrors could then allow arrays of identical single photon sources to be produced for potential uses in quantum information processing or used for studies involving strongly interacting polaritons in an optical lattice. 
\newline

This work has been supported by the EPSRC Programme Grant EP/J007544, ERC Advanced Investigator Grant EXCIPOL and by the Leverhulme Trust.

\bibliographystyle{aipnum4-1}

We have recently became aware of a preprint showing lateral polariton confinement using a fiber based approach.\cite{Besga2013}
\end{document}